\documentclass[twocolumn]{aastex61}

\usepackage[utf8]{inputenc} 
\usepackage{amsmath}

\newcommand{\sm}{\tilde{\sigma}/m_{\textrm{dm}}}
\newcommand{\rhodm}{\rho_{\textrm{dm}}}

\newcommand{\h}{\textrm{h}}
\newcommand{\msun}{\,\textrm{M}_{\odot}}
\newcommand{\unit}{\,\textrm{cm}^{2}\textrm{g}^{-1}}

\newcommand{\kpch}{\,\textrm{kpc}}
\newcommand{\msunpc}{\,\textrm{M}_{\odot}\textrm{pc}^{-3}}

\received{XXX}
\revised{YYY}
\accepted{ZZZ}
\submitjournal{ApJ}

\shorttitle{Disk Warps and SIDM}
\shortauthors{L. F. Secco et al.}

\begin{document}

\title{Probing Self-Interacting Dark Matter With Disk Galaxies in Cluster Environments}

\correspondingauthor{Lucas F. Secco}
\email{lucasfr@sas.upenn.edu}

\author{Lucas F. Secco}
\author{Amanda Farah}
\author{Bhuvnesh Jain}
\affil{Center for Particle Cosmology, University of Pennsylvania, 209 S. 33rd St., Philadelphia, PA 19104, USA }

\author{Susmita Adhikari}
\affil{Department of Astronomy, University of Illinois at Urbana-Champaign, 1002 West Green Street, Urbana, IL 61801-3080 USA}
\affil{Kavli Institute for Particle Astrophysics and Cosmology, Stanford University, 452 Lomita Mall, Stanford, CA 94305, USA}
\affil{Department of Physics, Stanford University, 382 Via Pueblo Mall, Stanford, CA 94305, USA}
\affil{SLAC National Accelerator Laboratory, 2575 Sand Hill Road, Menlo Park, CA  94025, USA}

\author{Arka Banerjee}
\affil{Kavli Institute for Particle Astrophysics and Cosmology, Stanford University, 452 Lomita Mall, Stanford, CA 94305, USA}
\affil{Department of Physics, Stanford University, 382 Via Pueblo Mall, Stanford, CA 94305, USA}
\affil{SLAC National Accelerator Laboratory, 2575 Sand Hill Road, Menlo Park, CA  94025, USA}
\affil{Department of Physics, University of Illinois at Urbana-Champaign, 1110 West Green Street, Urbana, IL 61801-3080 USA}

\author{Neal Dalal}
\affil{Department of Astronomy, University of Illinois at Urbana-Champaign, 1002 West Green Street, Urbana, IL 61801-3080 USA}
\affil{Department of Physics, University of Illinois at Urbana-Champaign, 1110 West Green Street, Urbana, IL 61801-3080 USA}
\affiliation{Perimeter Institute for Theoretical Physics, 31 Caroline Street North, Waterloo, ON N2L 2Y5, Canada}
 
\begin{abstract}
Self-Interacting Dark Matter (SIDM) has long been proposed as a solution to small scale problems posed by standard Cold Dark Matter (CDM). We use numerical simulations to study the effect of dark matter interactions on the morphology of disk galaxies falling into galaxy clusters. The effective drag force on dark matter leads to offsets of the stellar disk with respect to the surrounding halo, causing distortions in the disk. For anisotropic scattering cross-sections of 0.5 and 1.0$\unit$, we show that potentially observable warps, asymmetries, and thickening of the disk occur in simulations. We discuss observational tests of SIDM with galaxy surveys and  more realistic simulations needed to obtain detailed predictions.
\end{abstract}


\section{Introduction}

The successful standard cosmological paradigm assumes that the dominant fraction of the matter contained in the universe is in the form of a non-luminous, nearly collisionless component called \textit{dark matter} (DM). Furthermore, the clustering of matter in cosmological scales is often interpreted as evidence in favor of it being \textit{cold}, i.e. it was  non-relativistic at the time of its decoupling from the thermal bath of the primordial universe. 

However, the observed structure on small astrophysical scales has been claimed to be in tension with the CDM predictions derived from high-resolution simulations (e.g \cite{NFW1997, Dubinski1991} for an incomplete list). There are several aspects to this possible tension. In particular, CDM-only simulations predict that: (1) dark matter halo densities scale as $\rhodm(r)\propto r^{-1}$ in the inner $\sim1$ kpc of galaxies and (2) the number of satellite halos orbiting a Milky Way sized halo is $\mathcal{O}(100-1000)$. The empirical facts that some halos seem to have flat cores ($\rhodm \propto r^0$) and that only $\mathcal{O}(10)$ satellites have been found around our galaxy led these mismatches to be  named the ``core-cusp problem" and the ``missing satellites problem" respectively \citep{Kravtsov2010,Bullock2017}. Proper modeling and implementation of baryonic physics into simulations \citep{Wetzel2016} and the correction of observational biases \citep{Kim2017} have been claimed to alleviate such small scale structure problems, though it is unclear whether they are fully resolved (\cite{Rocha2013, Chan2015} and references therein). 

\cite{SpergelSteinhardt2000} suggested that  a nonzero cross-section $\sm$ for self-interacting dark matter (SIDM) could also help alleviate those problems. In the simplest model of SIDM, $\sm$ is velocity-independent and large when compared to weak-force scales. Shortly after that initial work, stringent constraints were derived based on different observational predictions of SIDM, for instance the sphericity (as opposed to triaxiality) of SIDM halos \citep{Miralda-Escude2002} and their evaporation rates \citep{GnedinOstriker2001}. With the growing sophistication of computational simulations, many of the previously obtained constraints have been significantly relaxed \citep{Dave2001,Rocha2013,Peter2013,Kim2016}. In particular, simulations also showed that some level of velocity dependence of the cross-section is necessary to simultaneously explain the core sizes of dwarf galaxies and clusters \citep{Yoshida2000,Colin2002,Vogelsberger2012,Zavala2013,Vogelsberger2014,Kaplinghat2016}.

Another potential observational consequence of dark matter self-interactions is an offset between light and mass centroids in cluster mergers. The dominant macroscopic effect of self-interactions in this case is expected to be analogous to that of a fluid-like drag force. The key idea is that, besides through gravitation, baryons are insensitive to SIDM, so stars act as a nearly collisionless component, while halos are decelerated by a drag force arising from the momentum transfer of DM interactions. The best studied example of such merger is the Bullet Cluster \citep{Bullet,Bullet2}. Constraints on $\sm$ have been derived on the basis that the separation between its matter centroid (as inferred from weak-lensing mass maps) and gas centroid (from X-ray emission) are of a few tens of kpc, consistent with zero within about the 68\% confidence level. Based on the offsets within the Bullet Cluster, \cite{Markevitch2004}  found $\sm < 5.0 \unit$, while simulations by \cite{Randall2008} found $\sm < 1.25 \unit$. Under SIDM, halos are also subject to evaporation. The upper bounds derived from the survival of the cluster despite halo evaporation are slightly more stringent: $\sm < 1.0 \unit$ from \cite{Markevitch2004} and $\sm < 0.7 \unit$ from \cite{Randall2008}. Roughly consistent constraints were obtained from studies of offsets within the Abell 3827 cluster \citep{Kahlhoefer2015_Abell} and from cluster collision images from HST and Chandra \citep{HarveyMassey2015}. Interestingly, the offsets between member galaxies and dark matter measured in those cluster mergers seem to be consistent with standard CDM when systematic effects of projection and mischaracterization of the centroids are significant \citep{Robertson2017,Ng2017}. The differing constraints from evaporation and centroid offsets arise from the different assumptions on the microscopic behavior of DM interactions. This difference is explored in greater detail in Section \ref{drag_theory}. It is commonly asserted that the desired range of cross-sections necessary to explain the observed mass profiles of galaxies is around $\sm\sim 0.5 - 5.0 \unit$ (\cite{BigReview} and references therein), with the upper bound being already severely constrained. It is thus notable that the simplest SIDM models have \textit{not} been unambiguously ruled out to date.   
 
In this context, galaxy clusters provide especially interesting environmental conditions for the study of dark matter interactions. If a flat core is present, the dark matter densities at the inner regions of $\mathcal{O}(10^{14}-10^{15})\msun$ cluster halos can reach around $10^{-2}\msunpc$, or even as dense as $10^{0}\msunpc$ in analytical cuspy profiles, while fairly concentrated galactic halos reach around $10^{-1}\msunpc$ in the inner 1$\,\mathrm{kpc}$. 

These densities enhance the number of interactions between dark matter particles, leading to potentially observable effects resulting from any additional drag force acting on DM. In this work, we consider a disk galaxy in its galactic halo (hereafter \textit{subhalo}) falling into a galaxy cluster (hereafter the main, \textit{host} halo). While offsets between the luminous components of cluster mergers and their total matter centroids have been measured, the analogous measurement for a galactic halo would likely be highly dominated by noise, since the weak lensing signal of a single such halo is not accurate enough to determine its centroid. However, we show that \textit{indirect effects} on disk galaxy morphologies resulting from relatively small baryon-DM displacements could be potentially measurable.             

In this work, we employ numerical simulations to characterize two types of distortions caused on stellar disks in SIDM subhalos, when an effective drag is the dominating factor. The first distortion is the \textit{warping} of the galaxy disk -- a U-shaped bending along the direction of motion. A second, longer lasting effect is the \textit{enhanced thickening} of the disk once the warp mode decays. 
U-shaped warps are not common in standard CDM -- tidal interactions and kinematic processes usually lead to S-shaped distortions, and HI disks are known to exhibit prominent S-shaped warps \citep{Binney1992}. For example, in a study on 26 edge-on disk galaxies by \cite{Garcia-Ruiz2002}, 21 HI warps were found, only two of which were U-shaped. However, these two were both highly disturbed and strongly interacting with visible nearby companions. In what follows, we show that dark matter self-interactions may lead to a U-shaped warp that is not necessarily caused by close encounters with neighboring galaxies.

This paper is organized as follows. In Section \ref{num}, we initially describe how SIDM can lead to a drag force. In that same Section, we specify the astrophysical system of interest -- a galaxy subhalo falling in a cluster -- and then describe the numerical simulations used. We also provide an analytical estimate of the simulation results. In Section \ref{results}, we focus on SIDM cross-sections of $\sm=0.5$ -- $1.0\unit$ and measure the intensities of warps and the enhancement in the thickness of simulated galaxy disks. We also perform variations around the fiducial system and test our numerical approximations in that Section. We conclude in Section \ref{discussion} with a discussion of our results and the potential applicability to data sets.



\section{Analytical and Numerical Approach to Model SIDM}\label{num}

\subsection{SIDM as a Drag Force}\label{drag_theory}

For colliding DM halos, the microscopic nature of the interactions between the dark matter particles determine the dominant macroscopic signatures that will be observable in such systems. The two main macroscopic effects that have been considered in literature are the \textit{evaporation} rate of the smaller halo, and an effective \textit{drag} that each halo experiences as it moves through the environment of the other halo. For short range interactions with isotropic cross sections, immediate halo evaporation is the most dominant effect (e.g. \cite{Markevitch2004} and \cite{Kahlhoefer2014}). As a large fraction of collisions have a high momentum transfer, particles may be expelled from their host's potential, leading to subhalo evaporation. The observation of surviving low mass subhalos therefore puts a stringent upper bound on the cross-section of isotropic scattering, as well as the fraction of collisions which result directly in the expulsion of dark matter particles from the subhalo.

On the other hand, if we consider interactions where the cross section is velocity-independent and anisotropic - that is, there are many more small angle scatterings than there are large angle scatterings, frequent interactions are possible without completely disrupting the subhalo. 
In this scenario, individual interactions are usually unable to expel particles from the subhalo. However, there is a non-zero cumulative evaporation rate resulting from multiple interactions. As shown in \cite{Kahlhoefer2014}, frequent and anisotropic self-interactions also lead to an effective ``drag'' force, and the rate of deceleration due to the drag force is comparable to the rate of cumulative evaporation. For these interactions, therefore, the macroscopic effects of the drag force can be comparable to the effects coming from evaporation \citep{Kummer2017}.

To understand the origin of the drag, we consider a two-particle interaction in the center of mass (COM) frame. If the scattering angle in this frame is $\theta$, then in the direction parallel to the relative velocity of the two particles, the change in the velocity of one particle is given by
\begin{equation}
\delta v_\parallel = -v(1-\cos \theta) \, ,
\end{equation}
where $v$ is the initial velocity of the particle in the COM frame. For isotropic interactions, $\cos\theta$ is drawn from an uniform distribution between $-1$ and $1$. For anisotropic interactions of the form that we are interested in, $\cos \theta$ is drawn from a distribution that is peaked near $1$ and $-1$, and has very low probability for collisions where $\cos \theta \sim 0$. In particular, we use the same form of the differential cross section for the anisotropic interactions as was used in \citep{Kahlhoefer2014}:
\begin{equation}
\frac{{\rm d} \sigma}{{\rm d}\Omega} \propto \frac{1+\cos ^2 \theta}{1-\cos^2 \theta} \,.
\label{cross}
\end{equation}

If collisions are frequent, and if individual interactions only change the initial velocity by a small amount, we can integrate over all possible interaction angles to obtain the average change in the velocity in the parallel direction of a particle passing through a sea of other particles. In a time interval $dt$, the number of interactions is given by
\begin{equation}
d N = \frac {\rho} {m_{\rm dm}} (v \, dt) \frac{d\sigma}{d\Omega}d\Omega \,,
\end{equation}
where $d\sigma/d\Omega$ is the differential cross section, $\rho$ is the ambient density, and $m_{\rm dm}$ is the mass of the dark matter particle. Using this, we can write the total change in the parallel velocity as 
\begin{equation}
\label{dvpar}
d v_\parallel = - \frac{\rho\, v^2\, dt}{m_{\rm dm}} \int\frac{d\sigma}{d\Omega}(1-\cos\theta) d\Omega \,.
\end{equation}
Since we assume that the dark matter particles are indistinguishable, we follow \cite{Kahlhoefer2014}, and define the momentum transfer cross section $\sigma_T$ as 
\begin{equation}
\sigma_T = 4\pi \int_0^1 \frac{d\sigma}{d\Omega}(1-\cos\theta) \, d(\cos\theta) \,.
\end{equation}
Once again, this expression makes sense only when the overall direction and velocity change per particle in the time interval $dt$ is small compared to the incoming velocity. For our choice of the differential cross section, this is a safe approximation to make for sufficiently small time steps. The integral runs over the scattering angle $0 \leq \theta \leq \pi/2$ since, for indistinguishable particles, a scattering angle above $\pi/2$ is the same as an equivalent scattering angle below $\pi/2$ but with a relabeling of the two outgoing particles. Using the expression for $\sigma_T$, equation (\ref{dvpar}) reduces to 
\begin{equation}
d v_\parallel = - \frac{\rho\, v^2 \, dt}{2m_{\rm dm}} \sigma_T
\end{equation}
Therefore, this change in the parallel velocity can be written as an effective drag deceleration due to the anisotropic self-interactions as a particle moves through a sea of other dark matter particles, with 
\begin{equation}
\label{eq:drag_eq}
\frac{F_{\rm drag}}{m_{\rm dm}} = -\frac 1 2 \left(\frac {\sigma_T}{m_{\rm dm}}\right) \rho v^2\,.
\end{equation}
We note that for isotropic cross sections where ${\rm d}\sigma/{\rm d}\Omega$ is a constant, the momentum transfer cross section $\sigma_T$ and the total cross section $\tilde \sigma$ is given by $\tilde \sigma = 2\sigma_T$. Therefore, we can re-cast equation (\ref{eq:drag_eq}) in terms of $\tilde \sigma$ so that our results can be compared directly to results for isotropic self-interactions. This yields
\begin{equation}
\frac{F_{\rm drag}}{m_{\rm dm}} = - \frac 1 4 \left(\frac {\tilde{\sigma}}{m_{\rm dm}}\right) \rho v^2\,.\label{eq: drag force}
\end{equation}
We note that we assumed that the relative velocity between the two colliding halos, or the velocity of a subhalo falling into the host halo, is much larger than the velocity dispersion of the subhalo, and therefore the self-interactions between the subhalo and the particles of the host halo dominate over self-interactions between particles of the subhalo itself.

\subsection{The Cluster and Subhalo System}\label{sec: physical_system}

\begin{figure*}
\noindent \begin{centering}
\includegraphics[scale=0.80]{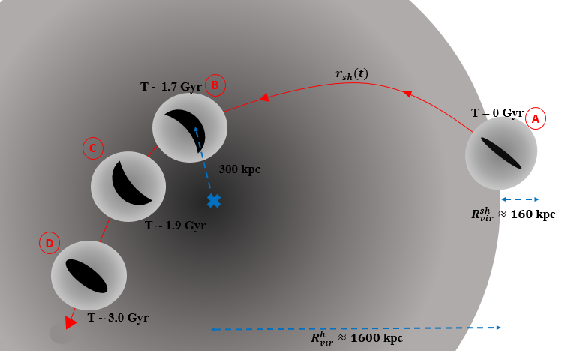}
\par\end{centering}
\caption{\label{fig: schematic} A schematic view of the simulated system (not to scale). With the fiducial choices detailed in Section \ref{sec: physical_system}, the virial radius of the host is around $R^\mathrm{h}_\mathrm{vir}=1600\kpch$ and that of the subhalo is around $R^\mathrm{sh}_\mathrm{vir}=160\kpch$. The curved red line corresponds to the trajectory of the subhalo though the host, $r_{\mathrm{sh}}(t)$. At (A), the thin galaxy disk is at the center of the subhalo at time $T=0\,\mathrm{Gy}$. At (B), at around $T=1.7\,\mathrm{Gy}$ (exact time values are slightly dependent on the chosen pericenter) the system is at closest approach from the host center and, shortly after that, the galaxy is maximally warped. At (C), typically  $200 \mathrm{My}$ after the initial forward warp, the distortion oscillates backwards. At much later times, (D), around $T=3\,\mathrm{Gy}$, the subhalo is close to its second turnaround or ``splashback radius" , and the disk is  thicker than it was at the start.}
\end{figure*}

We now describe the astrophysical system which is the focus of this study. Consider a spherically symmetric galactic subhalo which contains a disk galaxy seen edge-on by the observer. The centers of mass of both the subhalo and the disk initially coincide. The whole system moves along a trajectory $r_\mathrm{sh}$ with velocity $v_\mathrm{sh}$ through a host dark matter halo of density $\rho_{\textrm{h}}$ (a galaxy cluster), assumed to be at rest. For typical orbital velocities and halo densities, equation (\ref{eq: drag force}) leads to a non-negligible contribution of SIDM to the motion of the subhalo, decreasing its infall acceleration and consequently distorting the galaxy disk. We present a schematic description of the system in Figure \ref{fig: schematic}.

To simplify our treatment of this problem, we make several approximations
\begin{enumerate}
\item The galaxy disk is made of stars only, and they are perfectly collisionless. We assume that the  gravitational effect of the gaseous component of the disk is irrelevant to the dynamics.
\item The galaxy disk experiences the gravitational attraction towards the dark matter subhalo in which it is contained, but the subhalo does not experience any attraction towards the stars. While this approximation is used for most of the simulations in this work, we do test its validity in Section \ref{sec:numerical tests}.
\item The dark matter subhalo is spherical and is not distorted while crossing the host. It is characterized by a static radial density profile, so we treat its trajectory semi-analytically. The same applies to the host halo. This approximation implies that any possible disruption of either halos during the merger or evaporation of the subhalo has a small effect on the stellar disk; the effective drag force is dominant. 
\end{enumerate} 

With these approximations, plus equation (\ref{eq: drag force}), the acceleration of the subhalo center of mass and of a given star can be written as: 

\begin{align}
\ddot{\boldsymbol{r}}_{\textrm{sh}} & = -G\frac{M_{\textrm{h}}\left(<\left|\boldsymbol{r}_{\textrm{sh}}\right|\right)}{\left|\boldsymbol{r}_{\textrm{sh}}\right|^{3}}\boldsymbol{r}_{\textrm{sh}}+\frac{1}{4}\left(\frac{\tilde{\sigma}}{m_{\textrm{dm}}}\right)\rho_{\textrm{h}}\dot{\boldsymbol{r}}_{\textrm{sh}}^{2}\label{eq: on_subhalo}\\
\ddot{\boldsymbol{r}}_{\star}^{i} & = -G\frac{M_{\textrm{h}}\left(<\left|\boldsymbol{r}_{\textrm{\ensuremath{\star}}}^{i}\right|\right)}{\left|\boldsymbol{r}_{\star}^{i}\right|^{3}}\boldsymbol{r}_{\star}-Gm_{\star}\sum_{i\neq j}\frac{\left(\boldsymbol{r}_{\star}^{i}-\boldsymbol{r}_{\star}^{j}\right)}{\left|\boldsymbol{r}_{\star}^{i}-\boldsymbol{r}_{\star}^{j}\right|^{3}}\nonumber\\
 & -G\frac{M_{\textrm{sh}}\left(<\left|\boldsymbol{r}_{\star}^{i}-\boldsymbol{r}_{\textrm{sh}}\right|\right)}{\left|\boldsymbol{r}_{\star}^{i}-\boldsymbol{r}_{\textrm{sh}}\right|^{3}}\left(\boldsymbol{r}_{\star}^{i}-\boldsymbol{r}_{\textrm{sh}}\right),\label{eq: full_system}
\end{align}
where $\boldsymbol{r}_{\star}^{i}$ is the 3D position of $i$-th star relative to the center of the host, $M_{\textrm{h(sh)}}$ is the mass of the host (subhalo) enclosed in a given radius and $m_{\star}$ is the mass of a single star.

This system of equations is similar to that employed by \cite{Kahlhoefer2015_Abell}, to which we add the mutual gravitational attraction between stars by introducing the summation term of equation (\ref{eq: full_system}). In particular, disregarding that self-gravity term, direct numerical integration of equations (\ref{eq: on_subhalo}) and (\ref{eq: full_system}) has been used by those authors for an approximate description of the offset created between a colisionless component and its subhalo's center.

\begin{figure}
\noindent \begin{centering}
\includegraphics[scale=0.29]{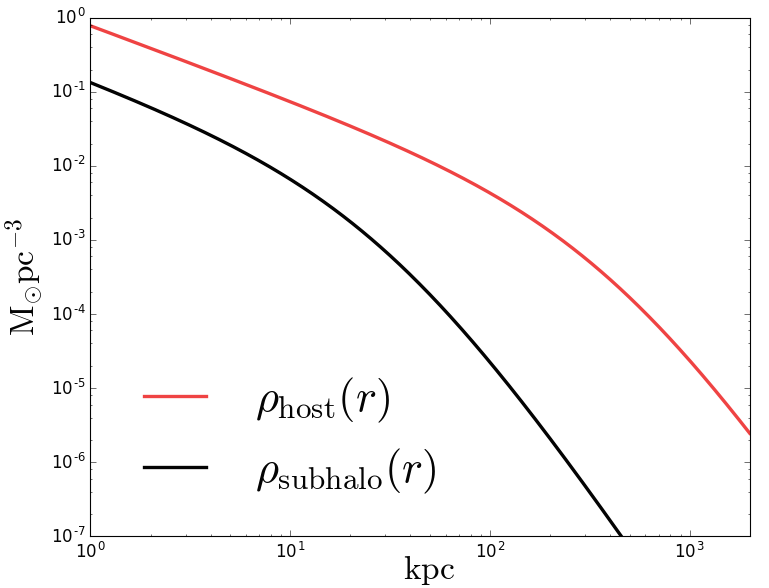}
\par\end{centering}
\caption{\label{fig: profiles}Radial density profiles of the subhalo (black) and the host halo (red). The host has a Hernquist profile defined by equation (\ref{eq: hernquist}), with virial mass $M_{\textrm{h}}=10^{15}\msun$ and concentration $c=5$. The subhalo is also a Hernquist profile with virial mass $M_{\textrm{sh}}=10^{12}\msun$ and concentration $c=8$. }
\end{figure}

Our choices for the halo profiles are shown in Figure \ref{fig: profiles}. For both the subhalo and host, we use a Hernquist profile \citep{Hernquist} given by 
\begin{equation}
\rho(r)=\frac{M_{\textrm{dm}}}{2\pi}\frac{a}{r(r+a)^{3}} \label{eq: hernquist}
\end{equation}
where $a$ is the scale factor and $M_{\textrm{dm}}$ is the total mass of the halo. Following \cite{SpringelMatteo2005}, we determine $a$ for each halo by matching their inner densities to an NFW profile \citep{NFW} with concentration $c$ and virial radius $r_{200}$ using
\begin{equation}
a=\frac{r_{200}}{c}\sqrt{2\left[\ln(1+c)-c/(1+c)\right]}.
\end{equation}
In our fiducial analysis, the host has halo mass of $10^{15}\msun$ and concentration $c=5$, while the subhalo has mass $10^{12}\msun$ and  $c=8$.

The disk galaxy contained in the subhalo has an exponentially decaying radial profile and a squared hyperbolic secant vertical profile:
\begin{equation}
\rho_{\textrm{disk}}(R,z)=\frac{M_{\textrm{disk}}}{4\pi z_{0}h_{0}^{2}}\exp\left(-\frac{R}{h_{0}}\right)\textrm{sech}^{2}\left(\frac{z}{z_{0}}\right)\label{eq: disk profile}
\end{equation}
where $h_0$ is the scale length of the disk, $z_0$ its scale height and $M_\textrm{disk}$ its stellar mass. We choose $M_{\textrm{disk}}=3\times 10^{10}\msun$, $h_{0}\approx 3.5 \kpch$ and a ratio of scale height to scale length of $z_0/h_0 = 0.1$ for the fiducial simulated galaxy disk. We also relate the velocity dispersions of the galaxy in the radial and vertical directions such that $\sigma_z = 0.5 \sigma_R$, and set $\sigma_\phi = \sigma_R$ for the azimuthal direction $\hat{\phi}$. With these choices, the resulting disk is roughly compatible with a Milky-Way type galaxy \citep{BinneyMerrifield1998}. The resulting galaxy is  stable and its  Toomre Q parameter is greater than 1 at all relevant radii \citep{ToomreQ}.


\subsection{Numerical Simulation}\label{sec: implementation_description}

To set up initial conditions for the phase space distribution of a galaxy disk, we use the publicly available code GalIC\footnote{https://www.h-its.org/tap-software-en/galic-code/} \citep{YurinSpringel2014}. GalIC iteratively minimizes a set of merit functions which measure the discrepancy between the target and realized spatial density field and velocity dispersions. It achieves this by using a Monte Carlo style algorithm that forces the velocity distributions to have the correct second order moments. We use $N=10^5$ particles to populate the stellar disk, and set the remaining parameters of GalIC to reproduce our fiducial scenario. The generated fiducial disk is close enough to an equilibrium state that it maintains its profile for a sufficiently long time.  

To evolve the motion of the subhalo-disk system in time, we need to efficiently solve equations (\ref{eq: on_subhalo}) and (\ref{eq: full_system}). We use the publicly available code GIZMO\footnote{http://www.tapir.caltech.edu/$\sim$phopkins/Site/GIZMO.html} \citep{GIZMO} in order to do so. GIZMO inherits some of its N-body algorithms from GADGET-2\footnote{http://wwwmpa.mpa-garching.mpg.de/gadget/} \citep{SpringelGADGET} and GADGET-3, and solves gravity with an efficient tree method. It also allows for the use of external analytic gravitational potentials and forces, and we use that feature to reproduce the motion of the subhalo through the host. Both halos are introduced as analytic potentials and are not populated with particles, and the subhalo moves according to a prescribed trajectory while the host is fixed. We first integrate equation (\ref{eq: on_subhalo}) with the relevant set of initial conditions and a choice of $\sm$ ($0.5\unit$, $1.0\unit$ or CDM), then feed to GIZMO a look-up table with this subhalo trajectory $r_\textrm{sh}(t)$.  
We describe in Section \ref{sec:numerical tests} tests in which both halos and galaxy are treated as collections of particles in the simulation and evolve under self-gravity, without analytic shortcuts, to confirm that this method is valid.   

As depicted in Figure \ref{fig: schematic}, the subhalo-plus-disk system starts its infall at the virial radius of the host, roughly $1600\kpch$ away from its center. 
To simulate realistic cases, we use subhalo orbits with several choices of pericenters, defined as the closest approach distances from the center of the host. We set the pericenter by initially giving the subhalo-disk system a tangential velocity whose magnitude is a fraction of the circular velocity about the host. While we experimented with different values, we focus on pericenter distances of 200 and 300$\,$kpc for our analysis, since these show appreciable SIDM effects and are not dominated by tidal distortions due to the host which could break our initial assumptions. The ratio of mean pericenter to cluster virial radius (apocenter) is about 1:6 as shown in simulations by \cite{Ghigna1998}.  We note that a more recent study by \cite{disruption2017} has found a mean pericenter that is larger by a factor of around two with respect to \cite{Ghigna1998}. Our choices thus reflect a sizable fraction of orbits of subhalos inside clusters. 

We also expect the orientation of the galaxy disk at its closest passage to the host to change the observable morphological distortions. This orientation is defined as the angle between the center of mass velocity of the galaxy disk and the normal to its plane. We study 4 different scenarios: orientation angles of 0, 22.5, 45 and 67.5 degrees (0 degrees is a face-on passing through the host and 90 degrees would be a passing along the diameter of the disk, which is not interesting for our analysis). Note that in all of these cases, the galaxy is still edge-on for the observer. In our simulations, each disk is evolved separately, so no simultaneous interaction with other galaxies is present. We describe the results for various choices of pericenters and orientation angles in Sections \ref{sec: warp_measure} and \ref{sec: puffy_measure}.


\subsection{An Analytical Prediction of the Results}\label{sec: analytic}

Consider a single star of mass $m_{\star}$ in a nearly circular orbit around the disk center. To determine the disk warping, we are interested in the star's motion in the direction perpendicular to the galaxy plane, $z(t)$. Under the approximation that the drag force is constant across the entire subhalo, and considering that the warp is small enough such that the disk potential is not severely disturbed, we can describe the motion of a star \textit{in the reference frame of the subhalo} as
\begin{equation}
\ddot{z}=\left(\frac{F_{\mathrm{drag}}}{m_{\mathrm{dm}}}\right)-\frac{F_{\mathrm{sh}}\sin\theta}{m_{\star}}-\frac{\partial\Phi_{\mathrm{eff}}(R,z)}{\partial z}\label{eq: sub_ref}
\end{equation}
where the first term is given by equation (\ref{eq: drag force}) and comes from the change in reference frames, and the effective potential of the axisymmetric stellar disk is $\Phi_{\mathrm{eff}}(R,z)=\Phi(R,z)+L_{z}^{2}/2R^{2}$, where $R$ is the radial distance from the star to the center of the disk along its plane, and $L_{z}$ its conserved angular momentum. The term $F_{\mathrm{sh}}\sin\theta$ is the projection of the subhalo gravity onto the direction perpendicular to the plane of the galaxy, with $\theta$ being the angle between the galaxy plane and the vector that goes from the galaxy center to the displaced star. One could also add to equation (\ref{eq: sub_ref}) a term that corresponds to the tidal force caused by the host, but that component is at least an order of magnitude weaker than the gravity of the subhalo at the scale radius of the disk.

Recalling the approximation that a star does not wander too far from the plane of the disk, we have $\sin\theta\approx z/R$. Using the spherical symmetry of the subhalo 
and the epicycle approximation \citep{B&T} to Taylor expand the potential up to second order in $z$-derivatives, we can write
\begin{equation}
\ddot{z}=\frac{1}{4}\left(\frac{\tilde{\sigma}}{m_{\mathrm{dm}}}\right)\rho_{\mathrm{h}}v_{\mathrm{sh}}^{2}-\frac{GM_{\mathrm{sh}}\left(R\right)}{R^{2}}\left(\frac{z}{R}\right)-\frac{\partial^{2}\Phi}{\partial z^{2}}z,\label{eq: pre HO}
\end{equation}
where the second derivative of the potential is evaluated at the equilibrium position $z=0$, on the plane of the galaxy. Describing the subhalo by a Hernquist profile, the cumulative mass within the radius $R$ is given by \citep{Hernquist}
\begin{equation}
M_{\mathrm{sh}}\left(R\right)=M_{\mathrm{dm}}\frac{R^{2}}{\left(R+a\right)^{2}},\label{eq: hernquist mass}
\end{equation}
where $a$ is the scale radius of the halo, equivalent to that of equation (\ref{eq: hernquist}). To treat the potential, we use a fully analytic expression that approximates the 3D potential of an axisymmetric disk with characteristic radius $h_0$ and characteristic height $z_0$, given by \cite{MiyamotoNagai}:
\begin{equation}
\Phi(R,z)=-\frac{GM_{\mathrm{disk}}}{\left[R^{2}+\left(h_0+\sqrt{z^{2}+z_0^{2}}\right)^{2}\right]^{1/2}}.\label{eq: MN}
\end{equation}

A final approximation is that most of the warp occurs in a relatively short time span, becoming maximal near the pericenter of the subhalo trajectory by the host. We verify with the simulations that this is especially true for the mild warps with which we are concerned in this derivation. With that in mind, the drag force acts as a ``kick'' on the star under consideration. Using equations (\ref{eq: hernquist mass}) and (\ref{eq: MN}) on equation (\ref{eq: pre HO}) and keeping the dominant orders in $R$, the motion of a point on the disk is described by
\begin{equation}
\ddot{z}(t=t_{\mathrm{imp}})=\left[\frac{1}{4}\left(\frac{\tilde{\sigma}}{m_{\mathrm{dm}}}\right)\rho_{\mathrm{h}}v_{\mathrm{sh}}^{2}\right]_{t=t_{\mathrm{imp}}}-\omega^{2}z\label{eq: HO}
\end{equation}
where $t_{\mathrm{imp}}$ is the time at which the subhalo is at pericenter, and where 
\begin{equation}
\omega^{2}=\frac{GM_{\mathrm{dm}}}{R\left(R+a\right)^{2}}+\frac{\left(1+\frac{h_0}{z_0}\right)R^{2}}{\left[(h_0+z_0)^{2}+R^{2}\right]^{5/2}}GM_{\mathrm{disk}}\label{eq: frequency}.
\end{equation}
such that the system resembles a driven harmonic oscillator with characteristic frequency $\omega$. 

Equation (\ref{eq: HO}) suggests two things: that the gravitational pull of the subhalo and disk act together as a restoring force, opposing the warp, and that this distortion is oscillatory. 
Especially at larger $R$, the squared frequency given by equation (\ref{eq: frequency}) goes as $R^{-3}$, so the restoring force is smaller at large radii and particles closer to the disk edge are less tightly bound gravitationally. 
The galaxy disk thus gets warped due to \textit{differential offsets} along the disk, its longitudinal section becoming ``U-shaped'', and the warp shape may oscillate around the initially thin configuration. Note that the edge-on disk initially bends \textit{forward} towards the direction of motion of the galaxy, as depicted on Figure \ref{fig: schematic}. 
We indeed find the warping effect in the simulations described in the following sections, not only for the mildly warped disks but also for the intense warps which perturb the galaxy potential beyond our analytic approximations.

We can use equation (\ref{eq: HO}) to obtain a theoretical estimate of the maximum amplitude of the warp. Consider the fact that very shortly after the drag-induced ``kick'', the galaxy reaches its new, distorted equilibrium position, and that the kick takes a finite, but small, amount of time. In that case, we can find the equilibrium displacement $z$ by setting $\ddot{z}=0$ on equation (\ref{eq: HO}). The maximal displacement obtained this way is:
\begin{equation}
z_{\mathrm{max}}\left(R\right)=\frac{\rho_{\mathrm{h}}v_{\mathrm{sh}}^{2}}{4\omega^{2}}\left(\frac{\tilde{\sigma}}{m_{\mathrm{dm}}}\right),
\label{eq: elegant_prediction}
\end{equation}
with $\omega^2$ given by equation (\ref{eq: frequency}). Using physical parameters similar to our fiducial system on equation (\ref{eq: elegant_prediction}) for the disk and subhalo dimensions and mass, we estimate the magnitude of the distortions in units of $\mathrm{kpc}$ on Table 1 for the pericenters of interest. More explicitly, we use $h_0=3.5\,\mathrm{kpc}$, $z_0=0.1\,h_0$, $M_\mathrm{dm}=10^{12}\msun$, $M_\mathrm{disk}=3\times10^{10}\msun$ and $a=30\,\mathrm{kpc}$. We use the local dark matter density of the host at pericenter, and the subhalo velocity is around $3200\,\mathrm{km/s}$.

\begin{table}
\begin{center}
\begin{tabular}{cccc}
 & $R=h_{0}$ & $R=2h_{0}$ & $R=3h_{0}$\tabularnewline
\hline 
$200\,\mathrm{kpc}$ & 0.48 & 1.05 & 2.39\tabularnewline
\hline 
$300\,\mathrm{kpc}$ & 0.19 & 0.42 & 0.95\tabularnewline
\hline 
\end{tabular}
\end{center}
\label{tab: prediction}
\caption{Analytic prediction of the magnitude of the displacement $z_\mathrm{max}$ as a function of disk radius, in units of $\mathrm{kpc}$, for the fiducial galaxy scenario when $\sm=1.0\unit$. These results match the simulations with reasonable accuracy.}
\end{table}
While equation (\ref{eq: elegant_prediction}) itself suggests a way of fitting the shape of the distortion, we choose a slightly different method in Section \ref{sec: warp_measure}, which accounts also for asymmetric warps. An added effect, not accounted for in the previous derivation, is an ``S-shaped'' warp resembling an integral sign. The presence of an S-shaped warp is a standard dynamical feature that arises quite generally from tidal interactions with close neighbors \citep{Binney1992,Reshetnikov1998,Shang1998}. 

Previous works have focused on the overall, center-of-mass offset between the dark matter halos and their luminous components, but overlooked the differential offset from an extended body such as a disk galaxy. In particular, while an offset between the centroids of light and total matter in cluster mergers must reach tens of kiloparsecs to be measurable, we find that displacements smaller than 1$\,$kpc between the subhalo and the stellar disk can lead to clear morphological distortions, as explored in Section \ref{sec: warp_measure}. 

However, perturbations that bend galaxy disks tend to damp out quickly, in a few rotation times \citep{B&T}, so the warps are not expected to be permanent features. Our simulations show that, after the warping phase, the stellar disk also does not revert back to a thin plane. Rather, in its final and relaxed state, sufficiently far from the densest parts of the host such that the drag force is again negligible, the stellar orbits spend significant time away from the galactic plane. We interpret this final state as an enhancement of the thickness of the initially thin disk. In Section \ref{sec: puffy_measure}, we quantify how much thicker the disk contained in the SIDM halo gets with respect to the baseline CDM case.


\section{Results}\label{results}   

\subsection{Measuring the Warping of Disks}\label{sec: warp_measure}

As expected from our analytical arguments, we indeed find that galaxy disks in our simulations warp forwards, then oscillate and warp backwards, and then finally relax in a puffier state than at the beginning. To quantify the intensity of the warps, we fit a 3rd order polynomial to the 2D projection of the edge-on stellar disk: $z(R) = aR^3 + bR^2 + cR + d$, where $z$ is the coordinate perpendicular to the plane of the disk and $R$ is the radial distance along the disk's plane, from its center.  Such a polynomial is able to capture the initial and final state of the disks, when it should be well fit by a straight line, as well as the U and S warps, which will force a fit to large $b$ and $a$ coefficients, respectively. This choice of metric is simple and convenient for our present purposes, but we note that others are available and have been used in the literature (e.g. \cite{Vikram2013}). To fit the polynomial when there is some orientation angle, we first rotate the galaxy so that it aligns with the vertical axis. Since the warps oscillate and then vanish, the fitting coefficients reproduce this behavior by getting increased in magnitude and then eventually decreased back to nullity. 
While the prediction from equation (\ref{eq: elegant_prediction}) assumed a symmetric warp and employed the approximation of a constant disk potential, which breaks down for strong warps, we do check that the predicted values in Table 1 are compatible with the simulations.

We take the magnitude of coefficient $b$ as a simple proxy of the U-warp intensity. To quantify some degree of theoretical uncertainty, we also obtain the error bars for $b$ from the covariance matrix of the polynomial fitting.  
For a distant galaxy, one would hardly observe stars significantly further than a few scale radii from the disk center, due to the rapidly decaying flux caused by the lower stellar density. To avoid overestimating the observable warp intensity, we restrict the polynomial fit only to particles that lie within a sphere of radius equal to $3h_0$ from the center of the subhalo, where $h_0$ is the scale radius of the galaxy disk.   

\begin{figure}
\noindent \begin{centering}
\includegraphics[scale=0.24]{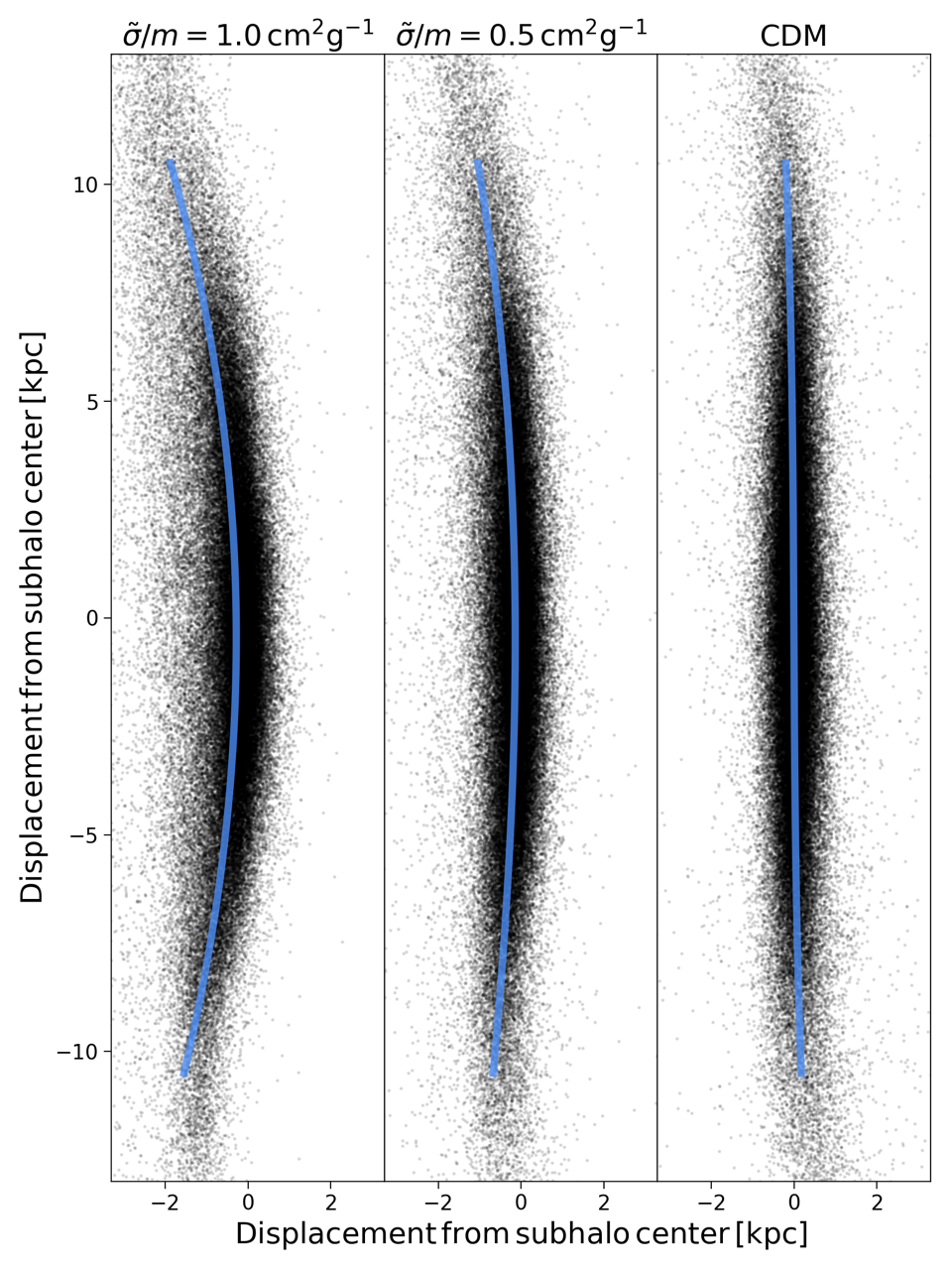}
\par\end{centering}
\caption{\label{fig: forward_warp}Snapshots from the simulations with 300$\,$kpc pericenter and orientation angle 0$^\circ$. Blue lines correspond to best-fit polynomials to the warp shape. From left to right, we display the two SIDM cross-sections analyzed and the standard CDM case. In all panels, the disk moves towards the left of the page, and the host's center is located towards the bottom of the page, as suggested by Figure \ref{fig: schematic}. The U-shaped warp is more severe on the case $\sm=1.0\unit$, and still very pronounced when $\sm=0.5\unit$, while the standard CDM case displays an S-shaped warp. The SIDM scenario is qualitatively very different than CDM.}
\end{figure}

\begin{figure}
\noindent \begin{centering}
\includegraphics[scale=0.24]{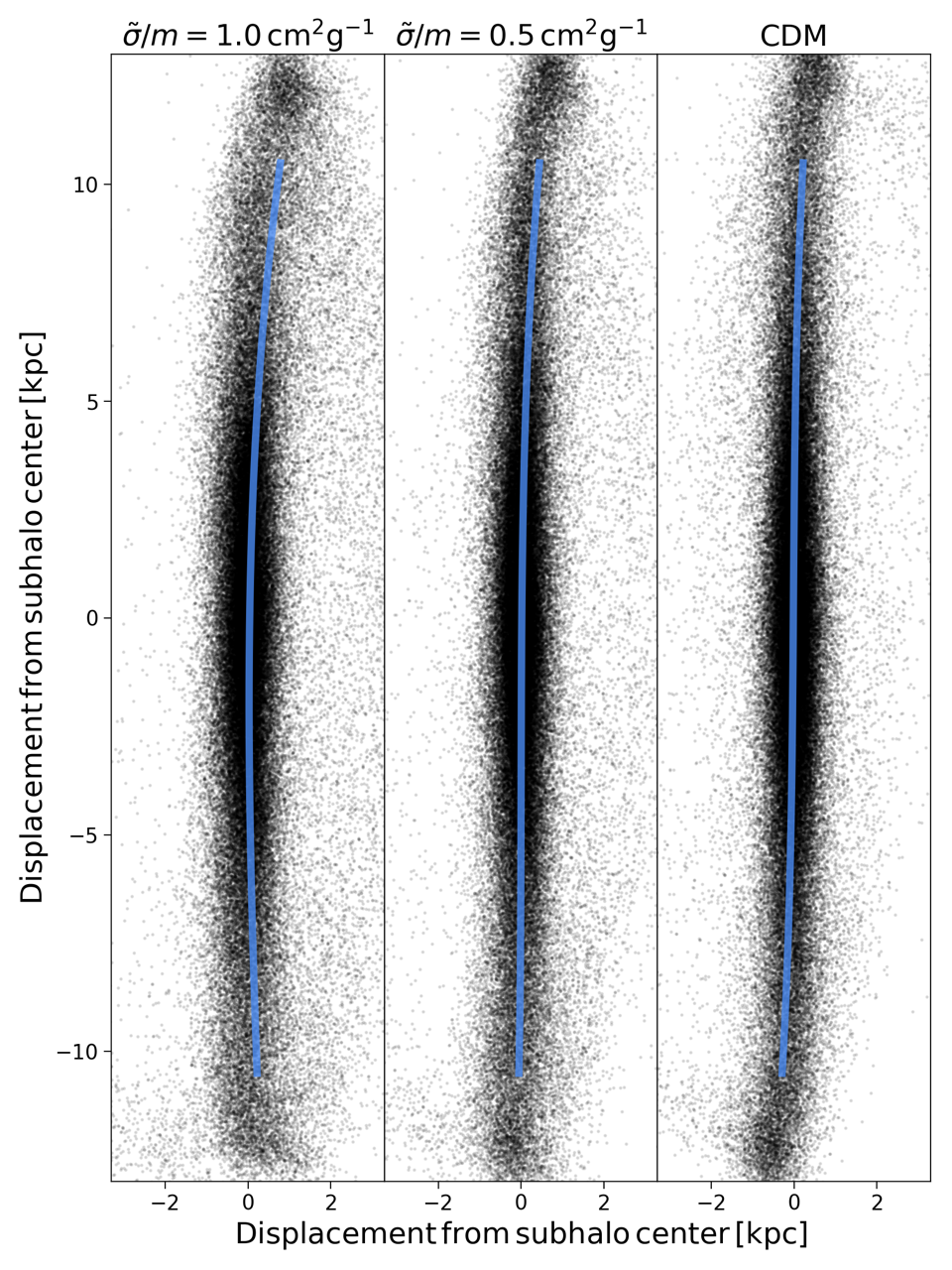}
\par\end{centering}
\caption{\label{fig: backward_warp}Same as Figure \ref{fig: forward_warp}, around 200$\,$My later. Both warps change orientation and the polynomial fit captures that change. As suggested again by Figure \ref{fig: schematic}, the host's center is located towards the right of the page.}
\end{figure}

For a better understanding of the distortions, Figures \ref{fig: forward_warp} and \ref{fig: backward_warp} 
show simulation snapshots of the evolution of the warps with time. With all stellar disks starting from the same initial conditions, we show in Figure \ref{fig: forward_warp} a snapshot of the $\sm = 0.5 \unit$ and $\sm=1.0 \unit$, 300$\,\mathrm{kpc}-$0$\,$deg simulation, as well as the CDM case. The galaxies in the SIDM halos exhibit prominent U-warps. Notably, the stellar disk that inhabits the CDM subhalo displays a modest S-warp at the same time frame. It is clear from the figure that the galaxy in the standard CDM subhalo should not display a significant $b$ coefficient as given by the polynomial fit. Notice that the $\sm=1.0\unit$ panel on Figure \ref{fig: forward_warp} provides a check for the analytical prediction of Table 1. There is overall good agreement between the values in the 300$\,\mathrm{kpc}$ row of that table and what is presented on Figure \ref{fig: forward_warp}. For $R=3h_0\gtrsim 10\,\mathrm{kpc}$, the predicted displacement of $0.95\kpch$ is in concordance with the visual displacement of around $1\kpch$ in that Figure.

Once the subhalo's gravity dominates over the drag again, the warp oscillates backwards, as represented by Figure \ref{fig: backward_warp} by the snapshots at a time step around 200$\,$My after those of Figure \ref{fig: forward_warp}. Similarly, the S-shaped warp on the galaxy belonging to the CDM subhalo changes orientation.

Finally, at a later stage, around 1$\,$Gy after the first warp and pericenter passage, the bending modes have decayed, and the remaining effect is an enhancement of the thickness of the stellar disk.  Such disk heating is also a feature of gravitational interactions in a standard CDM scenario. Our simulations suggest, however, that a population of disks under SIDM should be thicker than their CDM counterparts due to the extra heating caused by the initial warping. For a quantitative analysis, we determine a metric for the disk thickness and apply it to simulation snapshots on the following Section, where the thickness enhancement is made clear in Figure \ref{fig: hist_1}.

\begin{figure*}
\noindent \begin{centering}
\includegraphics[scale=0.48]{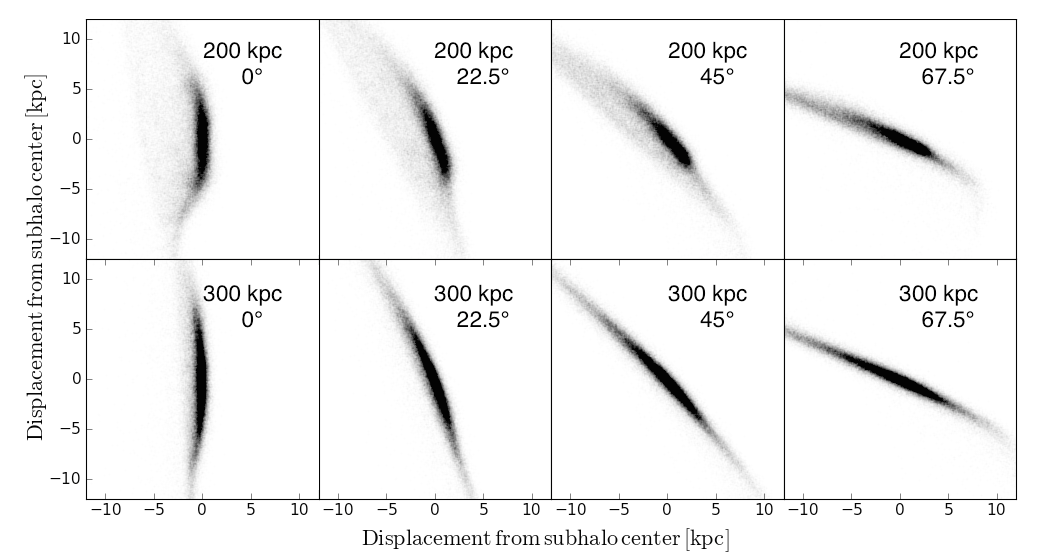}
\par\end{centering}
\caption{\label{fig: sm1.0}  Snapshots of the maximum forward warp in 8 simulations with different pericenters and orientation angles, for $\sm = 1.0 \unit$. Black dots correspond to individual stars on the disk that inhabits an SIDM subhalo. The same initial conditions were used for the galaxies on each panel. The U-shaped warp is more prominent for smaller pericenters (when the subhalo probes a higher ambient density $\rho_\h$) and lower orientation angles (when the collision is closer to face-on as seen by the host halo). We also note, but leave for a future study, the fact that there is a skewness in the light distribution of disk galaxies at orientation angles larger than 0 degrees. }
\end{figure*}

In Figure \ref{fig: sm1.0} we show snapshots of the simulations for both choices of pericenter and all orientation angles at the maximal warping instant according to the polynomial fit, for the case $\sm = 1.0\unit$. All panels show  galaxies that inhabit SIDM subhalos, and the disk initial conditions are the same. The units on each axis correspond to the separation of the edge-on disks to the center of their respective subhalos. The general trend is that the 0 degree orientation angle (face-on onto the host) leads to more prominent, symmetric warps, while other inclination angles are somewhat asymmetric around the center of the subhalo. Also, a shorter pericenter leads to stronger distortions as it probes denser parts of the host halo.

Figure \ref{fig: sm1.0} also shows that, for inclinations that are not exactly face-on (orientation angle differing from zero), the stellar distribution can become significantly skewed within the disk plane.  This skewness is another potential signature of DM self-interactions, in addition to the U-shaped warps and enhanced disk thickness.  Our simulations suggest that SIDM-induced skewness can persist within satellite disks for times similar to the durations of the U-shaped warps.  We defer a detailed study of SIDM-induced skewness to future work, and instead will focus on quantifying warp signatures and disk thickness.

\begin{figure*}
\noindent \begin{centering}
\includegraphics[scale=0.16]{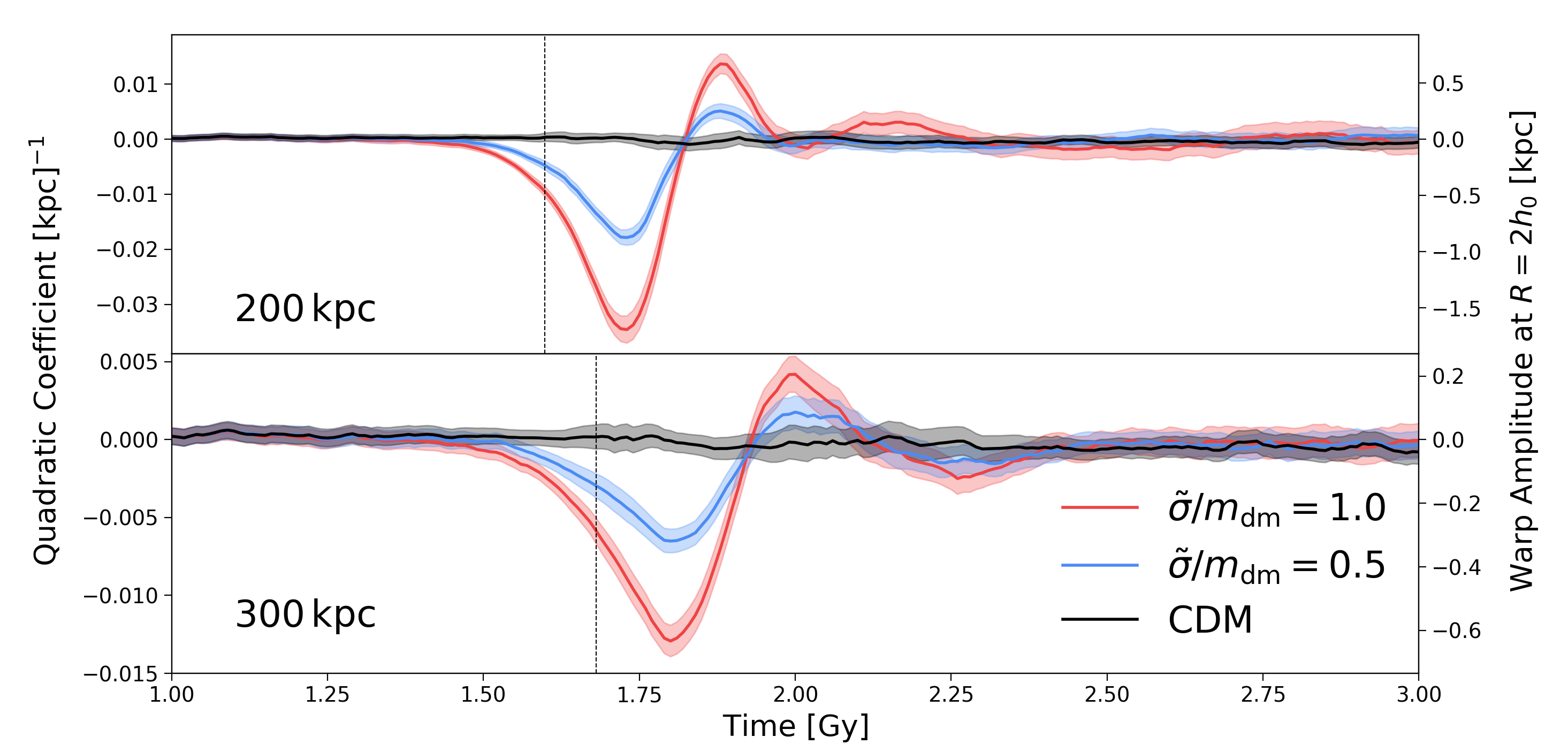}
\par\end{centering}
\caption{\label{fig: quadratic_coef} Quadratic coefficient $b$ of the polynomial fit $z(R) = aR^3 + bR^2 + cR + d$ as a function of time for two pericenters. This is a simple metric chosen among different possibilities (see text). On the right-hand axis, we show values for $z(R) = bR^2$ at $R=2h_0$ (the displacement of the stellar disk at 2 scale radii). Note that the quadratic coefficient alone may not reproduce the exact warps apparent on Figures \ref{fig: forward_warp} and \ref{fig: backward_warp}, where we fit the full polynomial. However, it captures the largest part of the observable warp effect. An estimate of the survival time of the warps is obtained from these curves (see text). Colored bands are estimates of the measurement error, obtained from the covariance of the fitted coefficient $b$, and are multiplied by a factor of 10 in the plot above for visualization. Vertical dashed lines mark the time at which galaxies reach the pericenter. Upper panel: 200$\,$kpc pericenter, 0$^\circ$ orientation angle. Lower panel: 300$\,$kpc pericenter, 0$^\circ$ orientation angle. In both cases, the CDM curve does not exhibit any appreciable U-shaped warp as given by this metric. }
\end{figure*}

In Figure \ref{fig: quadratic_coef} we show the evolution of the quadratic coefficient of the polynomial fit as a function of time for the  0$^\circ$ orientation angle in our 2 pericenters of interest. 
Negative values of $b$ correspond to a parabolic U-shape in the forward direction of motion, while positive $b$ values bend the disk backwards. In both panels, we see oscillations of the coefficient which eventually vanish as the disk aligns again with the subhalo center. The error bands are multiplied by a factor of 10 on those plots for better visualization.

We define the threshold for the detection of the warps to be $|b|>0.003$, which is roughly an order of magnitude larger than the values this same coefficient reaches in the baseline CDM case due to fluctuations. For a very symmetric warp with $z(R) \approx bR^2$, this threshold value of $b$ corresponds to a perpendicular displacement of approximately $0.03\,\mathrm{kpc}$ and $0.3\,\mathrm{kpc}$ as measured around the disk's scale radius and three times the scale radius, respectively. After the first backwards warp (the first positive bump of the quadratic coefficient $b$ on Figure \ref{fig: quadratic_coef}), the galaxy disk is significantly thicker than how is was initialized, looking ``puffy'' rather than warped. Conservatively restricting ourselves to only one forwards plus one backwards warp, we find from the curves on Figure \ref{fig: quadratic_coef} that the period during which the warp could be detected on the least affected 0-degrees orientation case ($\sm = 0.5 \unit$, 300$\,$kpc) is of around 0.1$\,$Gy, while the most affected disk ($\sm = 1.0 \unit$, 200$\,$kpc) is detectably warped for at least 0.4$\,$Gy. The survival time of the U-shaped warp ultimately translates into a potentially observable sample size, which we further discuss in Section \ref{discussion}.

An important difference then becomes clear: for SIDM cross-sections within the range that we have explored ($0.5 <\sm <1.0 \unit$), U-shaped stellar disks are expected to be quite general on galaxies that make a fast passing through a dense dark matter environment, such as a galaxy cluster. In standard CDM, such stellar disk warps should be rare when compared to their S-shaped counterparts. We expect that further work to measure the intensity of warps and how frequently they are found in observations can place tight constraints on $\sm$.

\begin{figure*}
\noindent \begin{centering}
\includegraphics[scale=0.18]{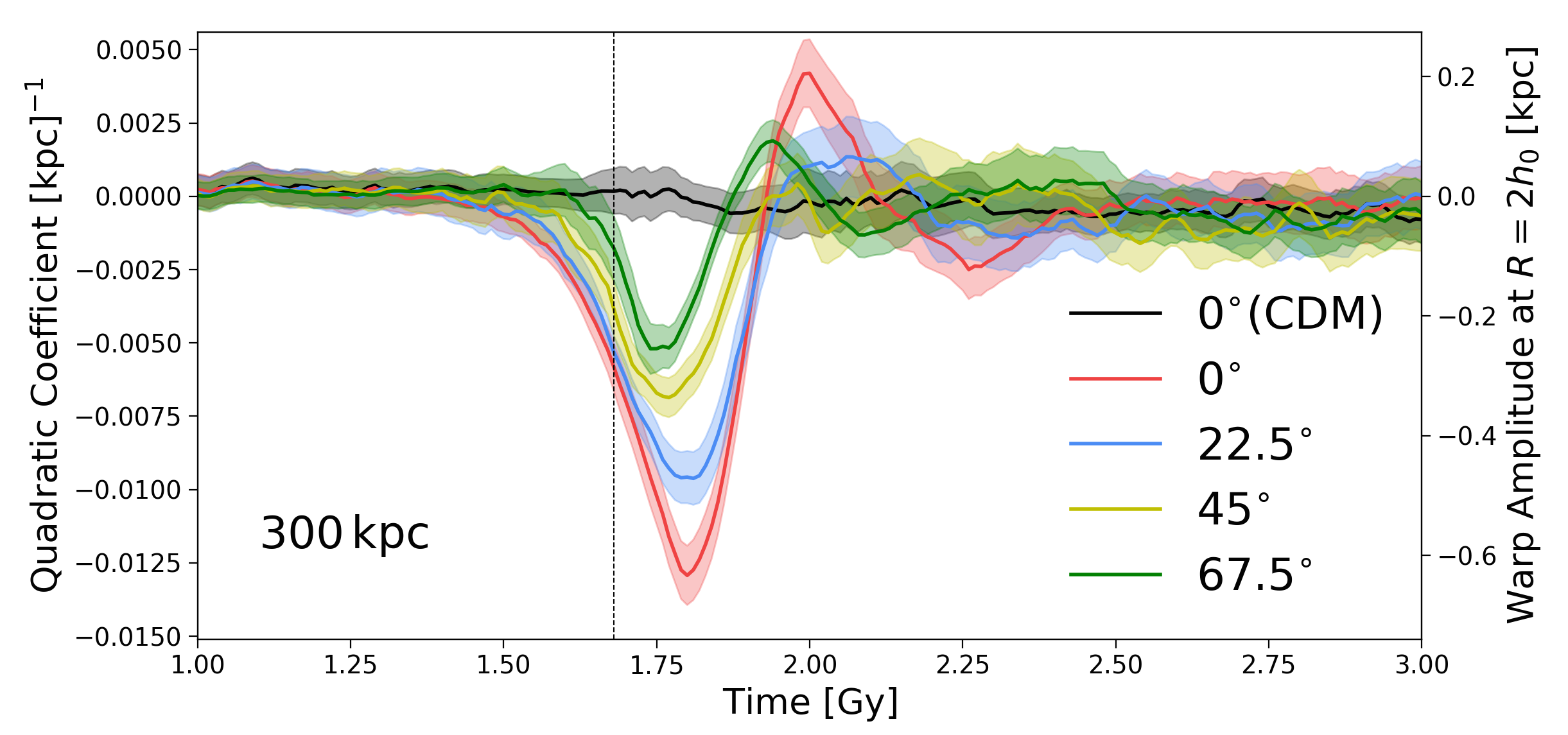}
\par\end{centering}
\caption{\label{fig: quadratic_coef_angles}  Same as Figure \ref{fig: quadratic_coef}, but for multiple orientation angles in the 300$\,$kpc pericenter case and cross-section $\sm = 1.0 \unit$. Steeper angles are not only more asymmetric (as seen from Figure \ref{fig: sm1.0}), but also display a less intense warp. Similarly to the 0$\,$deg case plotted above, other orientation angles for CDM are also null, as expected.}
\end{figure*}

To complement Figures \ref{fig: sm1.0} and \ref{fig: quadratic_coef}, we present Figure \ref{fig: quadratic_coef_angles} comparing different orientation angles for the case $\sm = 0.5\unit$ with 300$\,$kpc pericenter. We again show error bands which are multiplied by a factor of 10 for visualization. The trend is similar to that shown on Figure \ref{fig: sm1.0}, with steeper angles displaying a less intense -- but still detectable -- warp signal when compared to the symmetric 0$\,$deg case. 

\subsection{Measuring the Enhanced Disk Thickness}\label{sec: puffy_measure}

\begin{figure}
\noindent \begin{centering}
\includegraphics[scale=0.14]{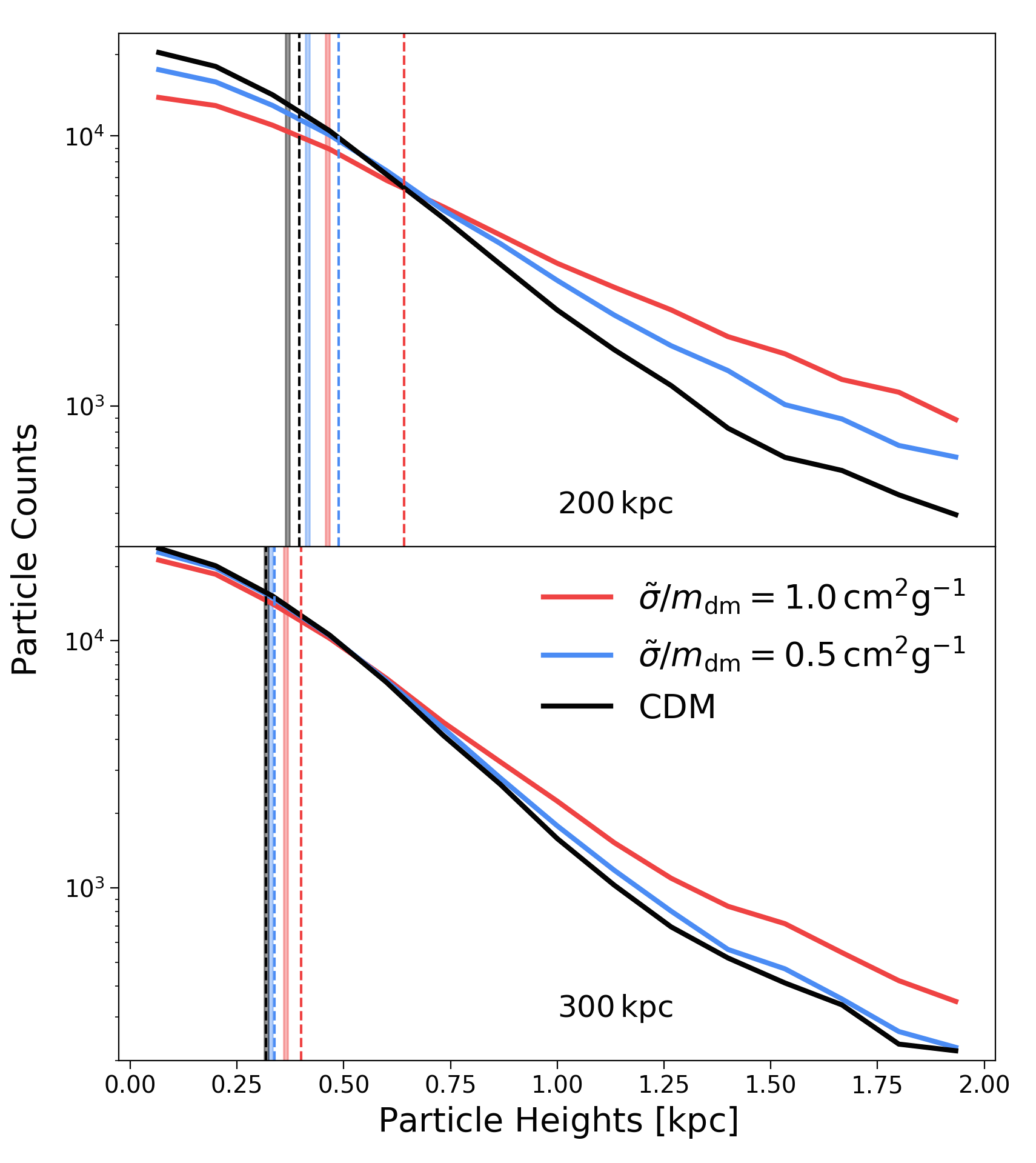}
\par\end{centering}
\caption{\label{fig: hist_1}  Histograms of disk particles selected to lie within a box of 3 scale radii and 5 scale heights from the galaxy center. Vertical bands correspond to the measurement of disk thickness as given by the standard deviation of the histograms, while dashed lines are scale height estimates from an exponential profile fit (see text). The uncertainty in the standard deviation metric is given by the band width and calculated from the mean $\Delta_{\mathrm{disk}}$ over 1$\,\mathrm{Gy}$. Upper panel: 200$\,$kpc pericenter, 0 degree inclination angle. The measured thickness is $\Delta_{\mathrm{disk}}(\mathrm{CDM})=0.32\pm0.006\,\mathrm{kpc}$, $\Delta_{\mathrm{disk}}(0.5\unit)=0.39\pm0.008\,\mathrm{kpc}$ and $\Delta_{\mathrm{disk}}(1.0\unit)=0.46\pm0.010\,\mathrm{kpc}$. The exponential fit yields scale heights $\tau(\mathrm{CDM})=0.34\,\mathrm{kpc}$, $\tau(0.5\unit)=0.49\,\mathrm{kpc}$ and $\tau(1.0\unit)=0.64\,\mathrm{kpc}$. Lower panel: 300$\,$kpc pericenter, also with 0 degree inclination angle. The enhancement with respect to CDM is only significant for the cross-section of $1.0\unit$.}
\end{figure}

Once the transient warps decay and the galaxy reaches a new equilibrium configuration, we find that the resulting stellar disks are thicker in the SIDM subhalos when compared to their standard CDM counterparts. This happens because stars in SIDM galaxies get scattered to outer orbits following the warped phase. To quantify the thickness enhancement, we use simulation snapshots at around 3$\,$Gy, when the disk is around its furthest distance from the host. We verify that the thickness is fairly constant around that time frame. On the 2D projection of the disk as seen by the observer, we first select only the particles that lie inside a box of radial range $R=[-3h_0,3h_0]$ and height range $z=[-5z_0,5z_0]$, where $h_0\approx 3.5\,\mathrm{kpc}$ is the initial scale length of the galaxy disk and $z_0=0.1h_0$ is the initial scale height. In the cases with some nonzero orientation angle, we first rotate the galaxies by the known angle to align them with the vertical axis, and then select particles. This cut is supposed to represent only particles that could be detected as part of the stellar disk in actual observations in the same way we select particles before fitting the 3rd order polynomial.

We then make histograms of the height of these selected particles with respect to the disk plane, which in the edge-on view of the observer is simply given by the horizontal coordinate $z$. As a primary metric for the disk thickness, we use the standard deviation of the resulting distribution of particles $\Delta_{\mathrm{disk}}\equiv\sqrt{\left\langle z^{2}\right\rangle -\left\langle z\right\rangle ^{2}}$. While the initial disk's transverse profile is, by construction, very well fit by a $\mathrm{sech}^2$ function according to equation (\ref{eq: disk profile}) from which a scale height can be obtained, we make this choice because the final, thicker disks are not well fit by that expression. 
As the thickness differences between CDM and SIDM play a more significant role at around $z\gtrsim3z_0$, we also fit an exponential profile proportional to $e^{-z/\tau}$ as a secondary metric, where $\tau$ is analogous to a new disk scale height. This fit is restricted within the range $[2z_0,4z_0]$, where the profiles are well described by a single exponential length scale $\tau$. For a comparison with a control disk, we also measure both thickness metrics in a simulated system that does not go through the host halo, but rather evolves in isolation in a CDM subhalo. After 3$\,$Gy, the vertical profile of this isolated disk remains consistent with the initial conditions due to GalIC's success in finding a quasi-equilibrium configuration. We find  $\Delta_{\textrm{disk}}^{\textrm{isol.}}=0.24\,$kpc and $\tau^{\textrm{isol.}}=0.20\,$kpc in this control system.

Figure \ref{fig: hist_1} represents the obtained histograms at time 3$\,$Gy for different cross-sections in the 200$\,$kpc and 300$\,$kpc pericenter cases, with orientation angle 0$^\circ$. 
The vertical bands on both panels show the values of the thickness metric, and the uncertainty represented by their width corresponds to an averaging over $\sim1\,\mathrm{Gy}$ around the turnaround time. For the 200$\,$kpc case we obtain $\Delta_{\mathrm{disk}}(\mathrm{CDM})=0.32\pm0.006\,\mathrm{kpc}$, $\Delta_{\mathrm{disk}}(\sm=0.5\unit)=0.39\pm0.008\,\mathrm{kpc}$ and $\Delta_{\mathrm{disk}}(\sm=1.0\unit)=0.46\pm0.010\,\mathrm{kpc}$. Notice that the CDM disk also gets thicker due to tidal interactions alone, which is a potential systematic uncertainty. On both panels, a larger cross-section leads to a thicker disk, but the discrepancy is more prominent for smaller pericenters, which probe a higher number of SIDM interactions, a trend similar to that of the U-shaped warps. Dashed lines correspond to the secondary metric, the exponential fit, which capture mostly the difference in the histogram tails. For the 200$\,$kpc case we obtain $\tau(\mathrm{CDM})=0.34\,\mathrm{kpc}$, $\tau(\sm=0.5\unit)=0.49\,\mathrm{kpc}$ and $\tau(\sm=1.0\unit)=0.64\,\mathrm{kpc}$. Both pericenters show a slightly stronger SIDM signal in the $\tau$ metric than in $\Delta_\mathrm{disk}$.

We note that including particles away from the limits $|z|<5z_0$ in the calculation of $\Delta_\mathrm{disk}$ increases the enhancement effect. That is mostly due to the fact that the disks in SIDM halos actually have a considerable number of particles spread out to much further heights, while the CDM histogram decays very quickly and has little contribution at $z>5z_0$, as can be seen from Figure \ref{fig: hist_1}. This implies that our selection of particles is conservative for observational purposes. 

The survival time for the enhanced thickness is considerably longer than that of the U-warps. We find that during at least 1 Gy the histograms presented in Figure \ref{fig: hist_1} remain very stable despite the orbital motion of particles. In fact, due to the collisionless nature of stars, it is not expected that the stellar disk will completely revert back to a thin plane, since that would require some energy to be radiated away. 

We thus conclude that this is another interesting difference between morphologies in SIDM and CDM in cluster environments. While, in both scenarios, a disk that has made a fast passage through a dense dark matter environment should become thicker, the currently allowed SIDM cross-sections can lead to an enhancement of that effect. A consequence of this conclusion is that field galaxies which have never been inside cluster environments should be relatively thinner, both in CDM and SIDM, than those inside clusters. 

A caveat is worth noting: one should expect the cluster environment to be more disturbing to disks than our single-galaxy simulations can capture. This could lead to a statistically increased  thickness of edge-on disks even in the absence of SIDM, for instance due to tidal interactions with many potential close neighbors other than the host itself.

\subsection{Variations of the Cluster and Galaxy Properties}\label{sec: tests}

As expected from our theoretical estimates in Section \ref{sec: analytic}, our results on the previous sections are somewhat sensitive to the initial choice of the fiducial system -- the host and subhalo masses and their concentration, the initial galaxy thickness, etc. We consider here variations of these physical parameters. In the next Section, we also vary the ``unphysical'' parameters of the simulation, for instance the number of particles and the gravitational softening scale. We also test, to some extent, the initial simplifying approximations made explicit in Section \ref{sec: physical_system}. To accomplish these tests, we simply run a new suite of simulations changing, whenever possible, only one relevant parameter at a time.

The physical characteristics of the system, like halo profiles and galaxy length scales, are ultimately a set of nuisance parameters that one would marginalize over, in some sense, when trying to infer the SIDM cross-section $\sm$ from actual observations. That is due to the fact that the results shown in Sections \ref{sec: warp_measure} and \ref{sec: puffy_measure} are expected to have a dependence on these parameters.

When we vary the host and subhalo profiles, the resulting warp strength follows the trend expected from the analytical derivation in Section \ref{sec: analytic}. Increasing the concentration of the subhalo, at fixed mass, from the fiducial choice $c=8$ to $c=12$ yields a more cuspy halo which consequently produces a stronger restoring force on the distorted galaxy disk. The disk warp is expected to be less intense. Indeed, the resulting warp curve for the 200$\,$kpc, $\sm=1.0\unit$ case then reaches a minimum coefficient $b=-0.024\,\mathrm{kpc}^{-1}$, approximately 30\% less intense than the fiducial case (see Figure \ref{fig: quadratic_coef}). The same trend is expected when the subhalo mass is increased from the fiducial $M_{\textrm{sh}}=10^{12}\msun$ to $5\times 10^{12}\msun$, and we confirm that the most intense warp coefficient is 50\% smaller than the fiducial scenario. Conversely, for a less massive subhalo with  $M_{\textrm{sh}}=5\times10^{11}\msun$, the maximum warp is 50\% larger than the fiducial case on Figure \ref{fig: quadratic_coef}. In these tests, the galaxy and host halo properties were kept the same as the fiducial choice.

We run another set of tests modifying the host profile parameters. We simultaneously change the initial infall distance of the disk galaxy to match the virial radius of the new host. The other galaxy parameters remain fixed at their fiducial values and again we look at the case 200$\,$kpc, $\sm=1.0\unit$ with 0$\,$deg orientation angle. Decreasing the host mass from $10^{15}\msun$ to $5\times10^{14}\msun$, at fixed concentration, reduces the maximum warp by around 60\%, resulting in $b=-0.014\,\mathrm{kpc}^{-1}$ at maximal warping. Also, further reducing it to $10^{14}\msun$ significantly reduces the maximum warp intensity to $b=-0.001\,\mathrm{(kpc)}^{-1}$, below our defined detection limit of $|b|>0.003\,\mathrm{(kpc)}^{-1}$. This again follows the trend described in Section \ref{sec: analytic}: reducing the mass at fixed concentration makes the host less dense at its center, leading to a weaker SIDM drag force from equation (\ref{eq: drag force}), and consequently subtler distortions.

We also modify the disk's structural parameters. Reducing the stellar mass from $3\times10^{10}\msun$ to $10^{10}\msun$ and changing the initial scale height from $z_0 = 0.1h_0$ to $z_0 = 0.05h_0$ and $z_0 = 0.2h_0$ did not produce changes larger than 10\% on the warp intensity. 

In all of the tests above regarding the physical parameters of the system, we also look at the effect on the disk thickening. With respect to our fiducial case, all variations caused only marginal changes on $\Delta_{\mathrm{disk}}$ below $\pm10\%$, to the final thickness of the stellar disk in the $\sm=1.0\unit$ SIDM subhalo (see Figure \ref{fig: hist_1}).


\subsection{Tests of Numerical Approximations}\label{sec:numerical tests}

We proceed to test the numerical, ``unphysical'' simulation parameters and our initial approximations. The (constant) softening length used in our simulation runs was of $50\,\mathrm{pc}$, considerably shorter than the length scales of the effects we have studied. We performed several runs with softening as small as $25\,\mathrm{pc}$ and as large as $200\,\mathrm{pc}$ and found no significant difference in the obtained results. We were focusing specially on the evolution of the disk thickness when this parameter changed, since spurious 2-body scatterings can lead to heating of the galaxy disk followed by an increase in its thickness.

To make sure our results are not an artifact of integration accuracy, we also experiment with GIZMO's tree construction frequency and the maximum allowed size of the time steps. We find no significant difference in our measurements when the tree updating is made faster by an order of magnitude, and the same for when the maximum time step allowed is made lower by one and two orders of magnitude. We also increase and decrease the number of disk particles by a factor of 2 and find subpercent differences in the warp intensity of thickness measurements. We thus conclude that our results are fully converged.

Finally, we test approximations (ii) and (iii) in Section \ref{sec: physical_system} by implementing simulations with both the subhalo and host described by dynamic, gravitating N-body particles. We run these N-body simulations by modifying GADGET-2 \citep{SpringelGADGET} to include self interacting dark matter with anisotroptic scattering cross-section \citep{BanerjeePrep}.  The disk was again generated using GalIC. The simulations used $2\times 10^7$ particles for the host halo of mass $10^{15}\msun$, $2\times10^4$ for the subhalo with mass $10^{12}\msun$ and $10^4$ particles in the $3\times10^{10}\msun$ disk. In general, anisotropic self-interactions will give rise to both evaporation of the subhalo as well as a drag. To isolate the effect of drag on the galaxy disk system we run a drag-only implementation where each simulation particle is considered to be representative of an ensemble of microscopic particles and the net drag force can be evaluated using equation (\ref{eq:drag_eq}). In this case we find that strong warps are produced in the  galaxy disk even when the outer regions of the subhalo are largely distorted by tidal interactions. This is consistent with our assumption that the warp is sensitive to the subhalo profile within the scales of the disk radius, which does not change significantly due to tidal interactions. The drag-only case by itself is not energy conserving. We run another set of simulations where we implement the full physical picture with drag and evaporation.  We treat interactions between simulation particles as if they represent actual microscopic particles - i.e. we use the total cross section $\sigma$ to decide if two neighboring particles interact or not, and then by choosing the scattering angle $\theta$  using the probability distribution for $\theta$ from the form of ${\rm d}\sigma/ {\rm d}\Omega$ \citep{BanerjeePrep}. Since these interactions change the velocities of the interacting particles in both the direction parallel to the relative velocity of the particles, as well as in the direction perpendicular to it, this method naturally incorporates the effects of both drag as well as subhalo evaporation arising from self-interactions.

Since evaporation affects the entire subhalo profile, we find that for the cross-section used in this work (equation \ref{cross}) the warping can be suppressed in our fiducial 300 kpc pericenter case. Conversely, the disk thickening is enhanced due to mass loss from the subhalo. However, when performing a full simulation for the idealized scenario of a completely radial trajectory for the subhalo, we find measurable disk warps, even though their magnitude is smaller than the drag-only case. It must be noted that the amount of evaporation and drag can be affected differently as both depend on the nature of $d\sigma/d\Omega$ \citep{Kummer2017}. These simulations are currently in the process of development and it will be of interest to do a detailed analysis on how different cross-sections affect the warping of galaxy disks.

\section{Discussion}\label{discussion}

We have studied the impact of dark matter self-interactions on the morphology of disk galaxies in galaxy cluster environments. The effective drag force of SIDM causes an offset between the dark matter subhalo and stars of the disk galaxy. The restoring force from the dark matter then causes the stellar disk to be distorted. We use modified N-body simulations to model anisotropic, velocity-independent SIDM and focus on  the morphology of disk galaxies as they pass through a large galaxy cluster. In Figures \ref{fig: forward_warp}-\ref{fig: sm1.0} we show the distortions of an edge-on disk. Our quantitative results focus on the symmetric, U-shaped warp and thickening of the disk. For SIDM cross-sections of $0.5$ to $1 \unit$, we find that a disk galaxy with pericenter $\lesssim 300\,\mathrm{kpc}$ gets significantly warped, and the warp oscillates on a timescale of a few hundred million years (Figures \ref{fig: quadratic_coef} and \ref{fig: quadratic_coef_angles}) before decaying and leaving a thickened disk (Figure \ref{fig: hist_1}). Thus we have identified the warping and thickening as distinct signatures of SIDM; more generally asymmetries in the light distribution arise once the disk is offset from the dark matter. 

Several caveats apply to our simulation findings and to a detailed connection to observations. We show tests with full simulations in which the dark matter halos are populated with particles, and are therefore susceptible to evaporation, rather than having their motion approximated by the drag force. Detailed studies with such simulations are needed to obtain the full range of SIDM effects and make accurate observational predictions. In addition, our predictions are expected to apply robustly only to galaxies for which the dark matter dominates the gravity at least towards the outer parts of the disk. 

The prospects for observational detection of these SIDM signatures also hinge on a robust understanding of disk galaxies in standard CDM, so that comparative statements can be made. This generally requires inclusion of gas physics. Disk galaxies are expected to be quenched, and therefore redder in color, and tidally distorted near the cluster center -- detailed predictions are challenging to obtain as feedback processes and other gas physics remains uncertain. There are several observational challenges as well. Perhaps the biggest observational uncertainty is in the true 3-D distance of observed disk galaxies from the cluster center. Other possible sources of uncertainty are the inclination angle with respect to the observer (when the disk is not exactly edge-on), the location of the cluster center and the dark matter density near it, and the gravitational influence of the galactic subhalo's dark matter over the stellar disk. The amount of time the disk galaxy has spent inside the cluster is an additional factor, as it may not be on its first passage. 

Given these caveats, one can take two approaches to connect robust SIDM predictions to observations. The first is to identify warped or otherwise distorted galaxies and compare their properties to CDM and SIDM predictions. The fraction of warped galaxies, the qualitative signature of the warp (S-shaped vs. U-shaped), and other features differ in the two models. Even a small sample of warped disk galaxies near cluster centers would be useful: the presence of U-shaped warps in such galaxies could support an SIDM explanation. This approach can already be implemented with more detailed predictions, as imaging of nearby disk galaxies with measured morphological properties has been obtained by the Sloan Digital Sky Survey (SDSS)\footnote{http://www.sdss.org/} and other  surveys. However, the lack of such warped disks would be harder to interpret:  if we only have projected positions, one must take into account the fact that, even in standard CDM, disk galaxies may simply not survive very close to the center, so  disk galaxies observed near the cluster center may have a large 3-D distance. 
The second approach is more statistical: analyze all disk galaxies as a function of projected distance from the cluster and compare to an ensemble of such galaxies in simulations. The presence of warps that last a few hundred million years suggest that a non-negligible fraction of disks at small radii would be observed to have U-warps in SIDM. 

High resolution, multi-color images of galaxies in galaxy clusters would be the appropriate sample for such tests. With ongoing and planned imaging surveys such as the Dark Energy Survey (DES)\footnote{https://www.darkenergysurvey.org/}, Subaru-HSC\footnote{http://hsc.mtk.nao.ac.jp/ssp/survey/}, the Kilo Degree Survey (KiDS)\footnote{http://kids.strw.leidenuniv.nl/}, the Wide Field InfraRed
Survey Telescope (WFIRST)\footnote{https://wfirst.gsfc.nasa.gov/} and Euclid\footnote{http://sci.esa.int/euclid/}, large samples of low-redshift clusters with well resolved galaxies will be available. In addition to the warping, thickening and other observed effects in simulations, one could also use velocity signatures from spatially resolved spectroscopy of face-on disks. We find that in the oscillation phase, beyond the scale radius the disk moves with an azimuthally symmetric velocity of 10's of km/s. This could lead to Doppler shifts between the edges and the center of the face-on disk. Yet another potentially interesting effect that we defer to future work is a skewness in the light profile of galaxies, as suggested by Figure \ref{fig: sm1.0}. In such a scenario, nearby elliptical galaxies would also be interesting candidates to probe SIDM cross-sections.   

A more detailed discussion of observational strategy is beyond the scope of this work since we have only simulated a few simple cases of disk galaxy infall. We leave for future work with detailed simulations the full prediction of galaxy morphologies in SIDM, and a more comprehensive connection to observational prospects. 


\section*{Acknowledgements}

We thank Philip Hopkins, Denis Yurin and Volker Springel for making their respective codes GIZMO, GalIC and GADGET-2 publicly available. We also thank Denis Yurin for clarifying GalIC's documentation in private correspondence. We are grateful to Tara da Cunha for help with running the simulations and helpful discussions. We also thank Raul Angulo, Eric Baxter, Jo Bovy, Vinicius Busti, Mariana Carrillo-Gonzalez, Basudeb Dasgupta,  Mike Jarvis, Andrey Kravtsov, Vinicius Miranda, Riccardo Penco, Eduardo Rozo, Jerry Sellwood, and Jake VanderPlas for useful comments and suggestions. LFS and BJ are supported in part by the US Department of Energy grant DE-SC0007901.    

\bibliography{sidm_bib}




\end{document}